\input{aipcheck}

\documentclass[
    ,final            
  ]
  {aipproc}

\layoutstyle{8x11single}
\newcommand{\Tcc}{T^\chi_c}
\usepackage{mathrsfs,amsmath,amssymb,amsfonts}

\begin{document}
\title{How the Polyakov loop and the  regularization affect strangeness and restoration of symmetries at finite $T$ }


\author{M. C. Ruivo}{
address={Centro de Física Computacional \\ 
Departamento de Física, Universidade de Coimbra,Portugal
}
}

\author{P. Costa}{
address={Centro de Física Computacional \\ 
Departamento de Física, Universidade de Coimbra,Portugal
}
}
\author{C. A. de Sousa}
{address={Centro de Física Computacional \\ 
Departamento de Física, Universidade de Coimbra,Portugal
}
}
\author{H. Hansen}{
  address={	Univ.Lyon/UCBL, CNRS/IN2P3, IPNL, 69622
	Villeurbanne Cedex, France}
 }

\author{W. M. Alberico}{
  address={Dipartimento di Fisica Teorica, University of
Torino and INFN, Sezione di Torino, via P. Giuria 1, I-10125 Torino, Italy}
}

\begin{abstract}
The effects of  the Polyakov loop and of a regularization procedure  that allows the presence of high momentum quark states at finite temperature is investigated within the Polyakov-loop extended Nambu-Jona-Lasinio model. The characteristic temperatures, as well as the behavior of observables that signal  deconfinement and restoration of chiral and axial symmetries, are analyzed, paying special attention to the behavior of strangeness degrees of freedom. We observe that the cumulative effects of the Polyakov loop and of the regularization procedure contribute to    a better description of the thermodynamics, as compared with lattice estimations.  We find a faster partial restoration of chiral symmetry and the restoration of the axial symmetry appears as a natural consequence of the full recovering of the chiral symmetry that was dynamically broken. These results show the relevance of the effects of the interplay among  the Polyakov loop dynamics, the high momentum quark sates and the restoration of the chiral and axial  symmetries at finite temperature. 

\end{abstract}

\maketitle


\section{Introduction}

The study of the QCD phase diagram in the $(T-\mu)$-plane and the search for signatures of the quark-gluon plasma have attracted an intensive investigation over the last decades. The output of this research is expected to play an important role in our understanding of the evolution of the early universe and of the physics of heavy-ion collisions at the BNL and at LHC (CERN). 

Phase transitions associated to deconfinement and restoration of chiral and axial U$_A$(1) symmetries are expected to occur at high density and/or temperature and the investigation of observables that signal such critical behaviors is a challenging problem. As an approach complementary to first-principle lattice simulations, we consider an effective model that can treat both the chiral and the deconfinement phase transitions, the Polyakov loop extended Nambu--Jona-Lasinio (PNJL) model \cite{Meisinger,Pisarski,Ratti,Hansen,Costa:2009}. This  model has the advantage of allowing  to interpret fairly the lattice QCD results and to extrapolate into regions not yet accessible to lattice simulations. 

A non trivial question in NJL type models is the choice of the regularization procedure. In fact, for some integrals the three dimensional cutoff is only necessary at zero temperature, the dropping of this cutoff at finite T allows for the presence of high momentum quark states, leading to interesting physical consequences, as it has been shown in \cite{Costa:2008}, where the advantages and drawbacks of this regularization have been discussed, in the framework of the NJL model. We will enlarge the use of this procedure to the PNJL model and discuss its influence on the behavior of several relevant observables. Let us notice that the choice of a regularization procedure is  part of the effective modeling of QCD thermodynamic. Indeed the presence of high momentum quarks is required to ensure that the entropy scales as $T^3$ at high temperature. 

Moreover, we shall investigate the role of the U$_A$(1) anomaly which, as is well known, is responsible for the flavor mixing effect that removes the degeneracy among several mesons and for the non zero value of the topological susceptibility, hence the restoration of axial symmetry should have relevant consequences on the behavior of these observables. The implementation of the new cutoff procedure in the NJL model \cite{Costa:2008} lowers the critical temperature for the phase transition, and shows that restoration of chiral and axial symmetries can also be a phenomenon relevant in the strange sector. In fact, the dynamically broken chiral symmetry is completely recovered, in both the strange and non strange sectors, leading to the restoration of the axial symmetry at about the same temperature: the quark masses go to the current values, the quark condensates and topological susceptibility vanish, the mixing angles go to the ideal values and the U$_A$(1) chiral partners  converge. 

On the other hand, we found \cite{Costa:2009} that the presence of the Polyakov loop brings the critical temperature to a better agreement with the most recent results of lattice calculations,  and deconfinement and  partial restoration of chiral symmetry  occur at very close temperatures. The restoration of chiral symmetry is  faster in the PNJL model than in the NJL one, leading to several meaningful effects: the meson-quark coupling constants show a remarkable difference in both models, there is a faster tendency to recover the Okubo-Zweig-Iizuka rule, and, finally, the topological susceptibility nicely reproduces the lattice results around $T/T_c \approx 1.0$. Moreover, due to the strange components of some mesons, the behavior of the strange quark mass (that decreases faster due to the Polyakov loop) is important for their properties, as well as for other observables related to the axial anomaly (as noticed in \cite{Costa:2005} the topological susceptibility is strongly influenced by the strange sector). 

The aim of this paper is to investigate the cumulative effects of the Polyakov loop  and of the infinite cutoff at finite temperature, having in mind to analyze the interplay among the U$_A$(1) anomaly, the Polyakov loop dynamics, the high momentum quark states and the restoration of chiral symmetry, paying special attention to the behavior of the strange quark degrees of freedom. In this concern, the characteristic temperatures, the spectrum of scalar and pseudoscalar meson chiral partners and the topological susceptibility will be analyzed.


\section{Model and formalism}

We perform our calculations in the framework of an extended  SU$_f$(3)
PNJL Lagrangian, which includes the 't Hooft instanton induced
interaction term that breaks the U$_A$(1) symmetry; moreover
quarks are coupled to a (spatially constant) temporal background gauge field
representing the Polyakov loop \cite{Fu:2007PRD,Ciminale:2007,Fukushima:2008PRD}:

\begin{eqnarray}
{\mathcal L_{PNJL}\,}&=& \bar q(i \gamma^\mu D_\mu-\hat m)q + \frac{1}{2}\,g_S\,\,\sum_{a=0}^8\, [\,{(\,\bar q\,\lambda^a\, q\,)}
^2\,\,+\,\,{(\,\bar q \,i\,\gamma_5\,\lambda^a\, q\,)}^2\,]\nonumber\\ & + & g_D\,\{\mbox{det}\,[\bar q\,(1+\gamma_5)\,q] +\mbox{det}
\,[\bar q\,(1-\gamma_5)\,q]\} - \mathcal{U}\left(\Phi[A],\bar\Phi[A];T\right). \label{eq:lag}
\end{eqnarray}
Here $q = (u,d,s)$ is the quark field with three flavors ($N_f=3$) and three colors
($N_c=3$), $\hat{m}=\mbox{diag}(m_u,m_d,m_s)$ is the current quark mass matrix and
$\lambda^a$ are the flavor SU$_f$(3) Gell--Mann matrices ($a = 0,1,\ldots , 8$), with ${
\lambda^0=\sqrt{\frac{2}{3}} \,  {\bf I}}$. The covariant derivative is defined as
$D^{\mu}=\partial^\mu-i A^\mu$, with $A^\mu=\delta^{\mu}_{0}A^0$ (Polyakov gauge); in
Euclidean notation $A^0 = -iA_4$.  The strong coupling constant $G_{Strong}$ is absorbed in the
definition of $A^\mu(x) = G_{Strong} {\cal A}^\mu_a(x)\frac{\lambda_a}{2}$, where ${\cal
A}^\mu_a$ is the (SU$_c$(3)) gauge field
and $\lambda_a$ are the (color) Gell--Mann matrices.

The  Polyakov loop field  $\Phi$   appearing in the potential term of
(\ref{eq:lag}) is related to the gauge field through the gauge covariant
average of the Polyakov line:
\begin{equation}
\Phi(\vec x)=<<l(\vec x)>>=\frac{1}{N_c}{\rm Tr}_c<<L(\vec x)>>,\,\,\,\,\,\,\,\,\,\,\,\mbox{with}\,\,\,\,\,\,L(\vec x) ={\cal P}\exp\left[i\int_0^\beta d\tau A_4(\vec x, \tau)\right]\,.
\label{eq:phi}
\end{equation}
The Polyakov loop is an order parameter for the restoration of the ${\mathbb Z}_3$ (the
center of SU$_c$(3)) symmetry of QCD and  is related to the deconfinement phase
transition: ${\mathbb Z}_3$ is broken in the deconfined phase ($\Phi \rightarrow 1$) and
restored in the confined one ($\Phi \rightarrow 0$).
Several  effective potentials for the field $\Phi$ are available in the literature. Here we use the following potential \cite{Ratti2}, which is known to give sensible results:

\begin{eqnarray} 
    \frac{\mathcal{U}\left(\Phi,\bar\Phi;T\right)}{T^4}
    =-\frac{a\left(T\right)}{2}\bar\Phi \Phi +
    b(T)\mbox{ln}[1-6\bar\Phi \Phi+4(\bar\Phi^3+ \Phi^3)-3(\bar\Phi \Phi)^2] 
\end{eqnarray}
with: 
\begin{eqnarray}     a\left(T\right)=a_0+a_1\left(\frac{T_0}{T}\right)+a_2\left(\frac{T_0}{T}
    \right)^2,\,\,\,\,b(T)=b_3\left(\frac{T_0}{T}\right)^3,
\end{eqnarray}
where: $a_0=3.51,\, a_1=-2.47,\, a_2=15.2,\,b_3= -1.75$ and $T_0=270$MeV. The parameter set used for the pure NJL sector is the same as in \cite{Costa:2005}.

Concerning the value of $T_0$, some comments are in order. Since one of the purposes of the present paper is to make a comparative study of  the effects of  two types of regularization,  we choose to plot the quantities under study on a relative temperature scale $T/T_c$, where $T_c$ is a characteristic temperature. As  noticed by \cite{Hansen} the dependence of the results on $T_0$ is mild 
and in this context the physical outcomes are not dramatically modified when 
one changes $T_0$. The  choice $T_0 = 270$ MeV appears 
to be the better one for this work because it ensures an almost exact 
coincidence between chiral crossover and deconfinement at zero chemical 
potential, as observed in lattice calculations.
The mean field equations are obtained by minimizing the thermodynamical potential $\Omega$, which now depends on Fermi-Dirac distribution functions modified  by the Polyakov loop,  with respect to the quark condensates 
$\left\langle \bar {q_i} q_i\right\rangle$ and to the fields $\Phi$ and $\bar \Phi$. The meson mass spectrum  is calculated using the same procedure described in \cite{Costa:2009} and references therein.


\section{Results and discussion}

We start our analysis by identifying the characteristic temperatures which separate the
different thermodynamic phases in the PNJL model \cite{Hansen}, with two regularization procedures at finite temperature: 
\begin{itemize}
  \item [(i)]{\em Regularization I}, where the cutoff is used  only in the integrals
that are divergent ($\Lambda \rightarrow \infty$ in the convergent ones).
\item [(ii)]{\em Regularization II},
where the  regularization consists in the use of the cutoff $\Lambda$ in all integrals.
\end{itemize}
\begin{figure}[h]
  	\includegraphics[width=.38\textheight]{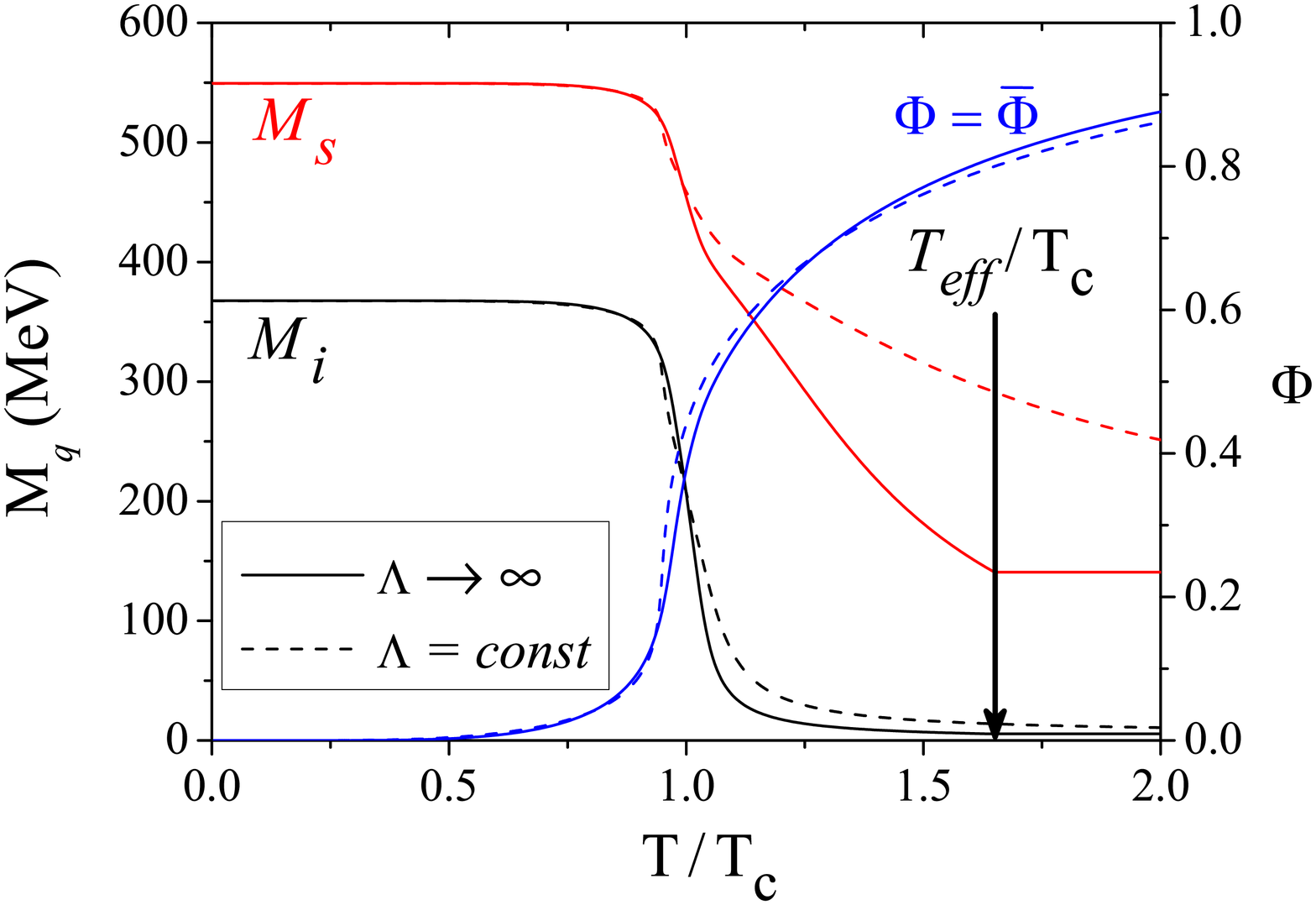}
    \includegraphics[width=.38\textheight]{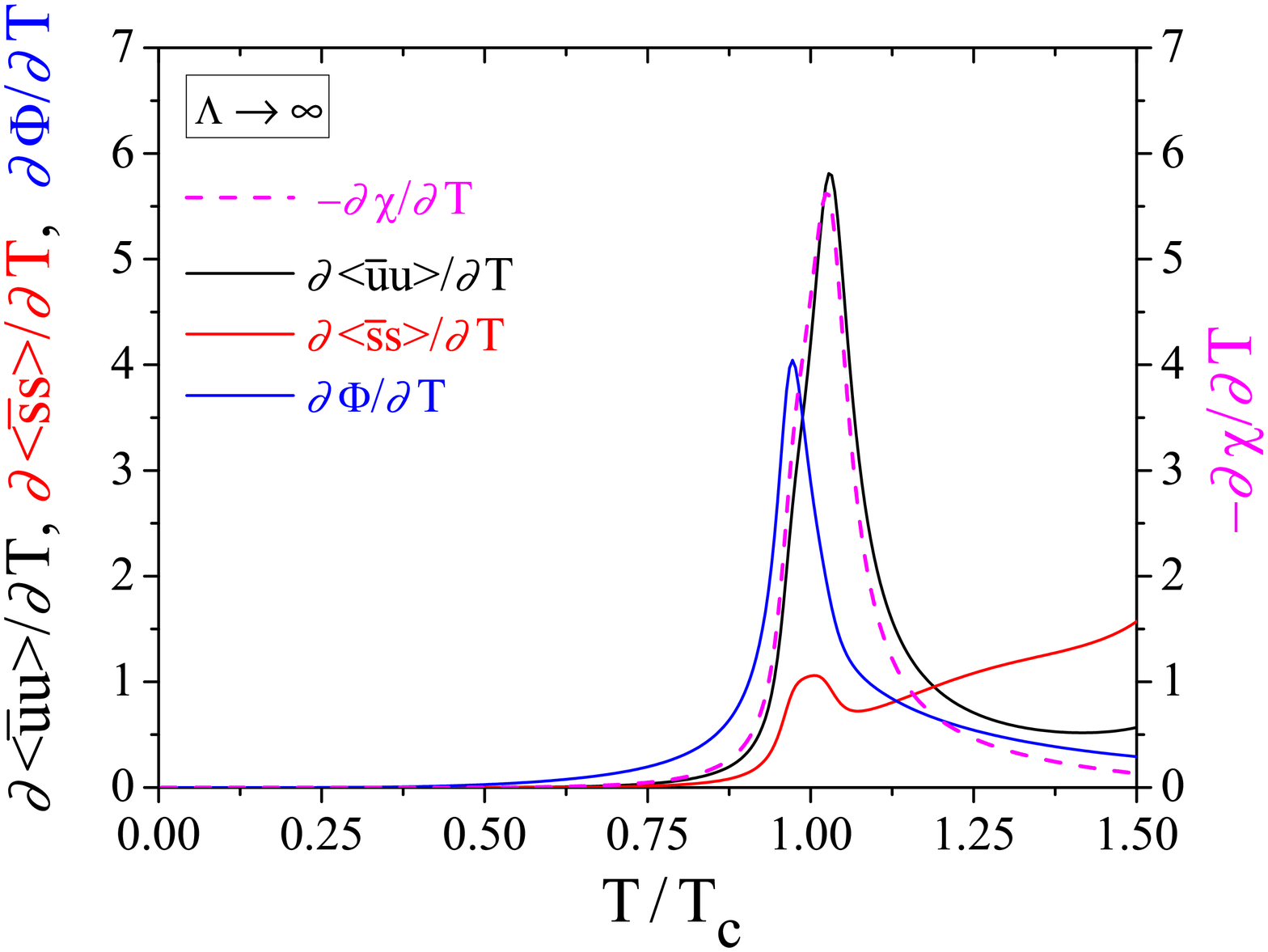}
		\caption{\label{fig:Mquarks} 
		Left panel: comparison of the quark masses 
		in PNJL model with regularization I (solid lines) and regularization II (dashed lines) as functions of the reduced 
		temperature $T/T_c$; the Polyakov loop crossover is also shown. At $T_{eff}=1.6T_c,  M_i=m_i$.
    Right panel: derivatives of the quark condensates and of the Polyakov loop field $\Phi$, with regularization I.} 
\end{figure}
\begin{table}[h]
\begin{tabular}{lccc}
\hline
& $T^\chi_c$  &  $T^\Phi_c$  &$T_c$   \\
& [MeV]& [MeV]&[MeV]\\
\hline
 I: $\Lambda \rightarrow \infty$ & 222 & 210 & 216 \\
\hline
 II: $\Lambda=const.$ &  258   &  234& 246\\
\hline
\end{tabular}
\caption{Characteristic temperatures.}
\label{table:paramNJL}
\end{table}
Let us analyze the comparative  effects of the two regularizations on the quark masses and the field $\Phi$. The critical temperature related to the
``deconfinement''\footnote{The terminology ``deconfinement'' in our model is used to
designate the transition between $\Phi \simeq 0$ and $\Phi \simeq 1$.} phase transition is
$T_c^\Phi$ and the
chiral phase transition characteristic temperature, $\Tcc$, signals partial restoration of chiral symmetry.    These temperatures are given, respectively,  by the inflexion points  of  $\Phi$ and of the chiral condensates  
$\langle\bar{q_i}q_i\rangle$,  and $T_c$ is defined as the average value between $T_c^\Phi$ and $\Tcc$. The main effect to be noticed is that  regularization I lowers the characteristic temperatures (see Table 1)  and decreases the gap $T_c^\Phi-\Tcc (24\mbox{MeV}\rightarrow 12\mbox{MeV})$, leading therefore to better agreement with lattice results. It is interesting to notice (Fig.\ref{fig:Mquarks}, right panel) that the inflexion points of $\Phi$ and of the strange quark condensate are almost coincident and slightly separated from the inflexion point of $\left\langle \bar{q_u} q_u\right\rangle$.
As already shown in \cite{Costa:2009}, the Polyakov loop leads to a faster decrease of the quark masses around $T_c$. Here we see that regularization I enhances this effect and at $T_{eff}=1.6 T_c$, it leads to the complete recovering of the chiral symmetry that was dynamically broken: the quark masses go to their current values and the quark condensates vanish. We remark that the effect of  allowing high momentum quark states is stronger for the strange quark mass. \footnote{We remark that  above $T_{eff}=1.6 T_c$ the quark masses become lower that their current values  and the quark condensates change sign. To avoid this unphysical effect  we use the approximation of imposing by hand the condition that, above $T_{eff}$, $M_i=m_i$ and  $\left\langle \bar{q_i} q_i\right\rangle=0$.} 

\begin{figure}[h] 
  \includegraphics[width=.45\textheight]{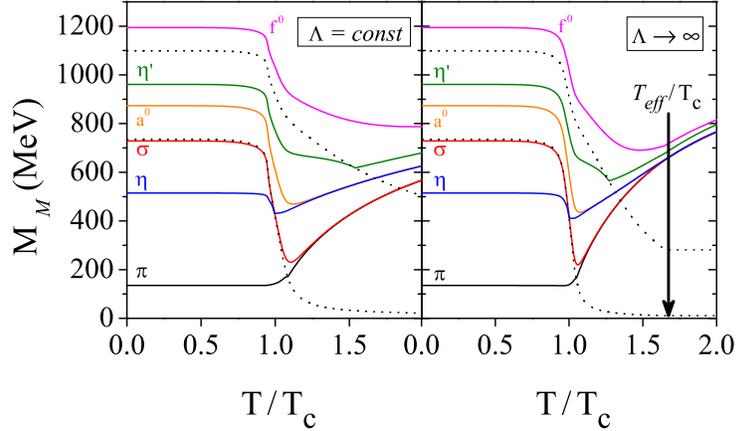}
   \caption{\label{fig:Mmesons} 
   Comparison of the pseudoscalar and scalar mesons masses in PNJL with regularization I  and regularization II   as functions
   of the reduced temperature $T/T_c$.}
\end{figure}

Since in our model chiral symmetry is explicitly broken by  non zero current quark masses, chiral symmetry is realized through parity doubling rather than by massless quarks. The effective restoration of chiral symmetry should be signaled by the degeneracy of meson chiral partners. In Fig.\ref{fig:Mmesons} we plot the masses of scalar and pseudoscalar mesons with the two regularizations. One notices that with regularization II the chiral partners $(\sigma, \pi)$ and $(a_0, \eta)$ degenerate, but not the chiral partners $(f_0, \eta')$. The analysis of the scalar and pseudoscalar mixing angles shows that they exhibit a tendency to go to their ideal values and that  $\sigma$ and $\eta$  mesons became less strange, while the $\eta'$ becomes more strange. The U$_A$(1) partners $(\pi, \eta)$ and $(a_0, \sigma)$ become close but do not converge. In conclusion, chiral symmetry is effectively restored in the non strange sector and axial symmetry is not restored. The  situation changes when regularization I is used: $(\sigma, \pi)$ and $(a_0, \eta)$ degenerate faster and $\sigma$ and $\eta$   became  purely non strange, while $(f_0, \eta')$ almost converge and $\eta'$ becomes purely strange. The U$_A$(1) chiral partners converge at $T_{eff}=1.6 T_c$ and the mixing angles attain the ideal values. So, chiral and axial symmetries are effectively restored both in the non strange and in the strange sectors.
\vskip0.1cm
\begin{table}[h]
\begin{tabular}{lcccc}
	\hline 
		$T_{Mott/T_c}$& $\sigma$  &  $\pi$  &$\eta$   \\
	\hline
		PNJL: $\Lambda \rightarrow \infty$ & 1.004 & 1.050 &1.004\\
	\hline
		PNJL: $\Lambda=const.$ &  0.963   &      1.085& 1.000\\
	\hline
		NJL: $\Lambda=const.$ & 0.820   &        1.086& 0.920\\
	\hline 
\end{tabular}
\caption{Mott temperatures.}
\label{table:MottPNJL}
\end{table}
\vskip0.1cm
As it is well known, at the Mott temperature the mesons dissociate into $q \bar q$ pairs and cease to be bound states. The Polyakov loop has the effect of increasing the Mott temperature (see \cite{Costa:2009}) of the mesons whose strangeness content decreases, like the $\sigma$ and the $\eta$. We found that  regularization I  enhances this effect (see Table 2), which means that these mesons survive as  bound states for temperatures slightly higher than the critical temperature.
\begin{figure}[h]
	\includegraphics[width=.45\textheight]{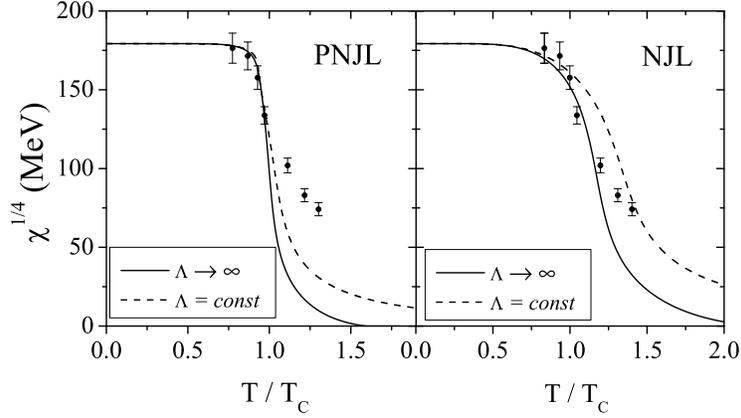}
	\caption{\label{fig:topo} 
	The topological susceptibility, as compared with lattice results, in PNJL  and NJL models, with two regularizations.} 
\end{figure}
Finally we analyze the topological susceptibility (Fig.\ref{fig:topo}). We notice that with both regularizations the topological susceptibility has a sharp decrease around $T_c$ and there is a nice adjustment to the first  lattice points. This effect is mainly due to the Polyakov loop, which is dominant around $T_c$; as the temperature increases   regularization I leads to a sharper decrease of the topological susceptibility, that vanishes at $T_{eff}=1.6T_c$, the temperature at which the dynamically broken chiral symmetry is completely restored.  A close connection between the restoration of chiral and axial symmetries is found. Indeed, the inflexion point of the topological susceptibility, $\chi$,   coincides with the one of  $\left\langle \bar{q_u} q_u\right\rangle$ and is very close to the one of $\left\langle \bar {q_s} q_s\right\rangle$   (see Fig.1, right panel) and the vanishing of these quantities occur simultaneously  at $T_{eff}=1.6 T_c$ (when  $M_i=m_i$). It can be noticed that the comparative effect of the two regularizations is qualitatively similar in PNJL and NJL model (Fig.\ref{fig:topo}, left panel). The Polyakov loop  fasten the effects observed in NJL model and leads to a nice fit to the first four lattice points.

In conclusion, we observe that, while the effects of the Polyakov loop are dominant around the critical temperature, the relevance of the presence of high moment quark states is seen also  at higher temperatures and influences mainly the strange sector. By combining both effects, we get a better description of the thermodynamics, as compared with lattice estimations; we find a faster partial restoration of chiral symmetry and the restoration of the axial symmetry appears as a natural consequence of the full recovering of the chiral symmetry that was dynamically broken. These results show the relevance of  the interplay among U$_A$(1) anomaly, the Polyakov loop dynamics, the high momentum quark states and the restoration of the chiral symmetry at finite temperature. 

\begin{theacknowledgments}
  Work supported by  project CERN/FP/83644/2008, FCT.
\end{theacknowledgments}

\bibliographystyle{aipproc}   
\bibliographystyle{aipproc}   


\IfFileExists{\jobname.bbl}{}
 {\typeout{}
  \typeout{******************************************}
  \typeout{** Please run "bibtex \jobname" to optain}
  \typeout{** the bibliography and then re-run LaTeX}
  \typeout{** twice to fix the references!}
  \typeout{******************************************}
  \typeout{}
 }

\end{document}